\documentclass[prb,aps,amsmath,preprint]{revtex4}
\usepackage{graphicx}
\usepackage{dcolumn}
\usepackage{bm}
\usepackage{subfigure}
\usepackage{scrextend}
\usepackage{hyperref}

\begin{document}

\title{High pressure structural, electronic, and optical properties of polymorphic InVO$_4$ Phases}
\author{S. Mondal}
\affiliation{School of Physics, University of Hyderabad, Prof. C. R. Rao Road, Gachibowli, Hyderabad-500046, India }

\author{S. Appalakondaiah and G. Vaitheeswaran$^*$}
\affiliation{Advanced Centre of Research in High Energy Materials (ACRHEM),\\
University of Hyderabad, Prof. C. R. Rao Road, Gachibowli, Hyderabad - 500 046, India.}
\date{\today}

\begin{abstract}
In the present work, we report a detailed density functional theory calculation on polymorphic InVO$_4$ phases by means of projector augmented wave method. The computed first-order structural phase transformation from orthorhombic \emph{(Cmcm)} to monoclinic \emph{(P2/c)} structure is found to occur around 5.6 GPa along with a large volume collapse of 16.6$\%$, which is consistent with previously reported experimental data. This transformation also leads to an increase in the coordination number of vanadium atom from 4 to 6. The computed equilibrium and high pressure structural properties of both InVO$_4$ phases, including unit cell parameters, equation of state, and bulk moduli, are in good agreement with the available experimental data. In addition, compressibility is found to be highly anisotropic and the \emph{b}-axis being more compressible than the other for both the structures. Electronic band structures for both the phases were calculated, and the band gap for orthorhombic and monoclinic InVO$_4$ are found to be 4.02 and 1.67 eV, respectively, within the Tran-Blaha Modified Becke-Johnson potential as implemented in linearized augmented planewave method. We further examined the optical properties such as dielectric function, refractive index, and absorption spectra for both the structures. From the implications of these results, it can be proposed that the high pressure InVO$_4$ phase can be more useful than orthorhombic phase for photo catalytic applications.

\end{abstract}

\maketitle

\section{Introduction}

ABO$_4$ type ternary oxides are very important from the theoretical and technological point of view.\cite{E1,E2,Solano,Fukunaga,Li,Oshikiri} The fundamental physical and chemical properties of these kinds of materials have been investigated at normal conditions and under compression through different experiments and theoretical calculations. For example, different types of tungstate (WO$_4$) are promising materials for the next generation cryogenic phonon-scintillation detectors, laser-host materials, optoelectonic devices, and mining.\cite{JR,Faure,Bohacek,Paski,Goel,Nikl,Annenkov,Kobayashi} Also, materials belonging to the orthovanadates (VO$_4$) have been widely investigated for their efficient luminescent properties, promising photocatalytic properties, low loss optical planar wave guides, capability to convert the ultraviolet emission into visible light, etc.\cite{Panchal,Oshikiri1,Paszkowicz,Ronde,Kaminskii,Fujimotoa, Voloshina,Chen,Shwetha} These investigations revealed that these compounds possess very attractive properties, which are in fact technologically important within a wider scope.\\
The present computed compound InVO$_4$ is one of the orthovanadate, which has attracting electrochemical and photocatalytic properties and can be used in different industrial applications.\cite{V1,V2,Liu}
Researches on InVO$_4$ have revealed that it can be used as photovoltaic cells for solar energy utilization and as a photocatalyst that is able to induce hydrolysis of water molecules under visible light irradiation.\cite{Oshikiri} It is also useful as electrolyte for lithium ion batteries since it undergoes conversion of indium, vanadium oxide, and alloying reactions of indium\cite{V3} and electrochromic windows due to their transparency.\cite{Vuk} The reason that makes its use as a good photovoltaic cell is the vanadium 3\emph{d} band, which is situated below the analogous \emph{d} band of the other transition metals in the energy spectrum decreasing the band gap.\cite{Oshikiri}\\
Previous studies on InVO$_4$ were limited to crystal structure, vibrational properties, and optical properties at high temperatures. These studies also included the photocatalytic properties of InVO$_4$ in bulk and in thin-film configurations.\cite{McCrone,Touboul1,Daniel,Baran,Yea,Enache,Katari,Oshikiri11} Recently, Errandonea et al.\cite{Daniel} experimentally reported a high pressure phase transformation of InVO$_4$ from CrVO$_4$ type (crystallizing in orthorhombic structure) to wolframite type (crystallizing in monoclinic structure) phase with a large volume collapse. The transition is associated with a change in coordination number around vanadium atom. From this, we can expect some changes in the electronic structure for the high pressure phase of InVO$_4$. Earlier experimental electronic properties of orthorhombic InVO$_4$ have been reported using optical absorption measurements.\cite{V2, Yea, Enache} Band gap values are around 2 eV for bulk structure and 3.5 eV for thin film (at nearly 723 K). On the other hand, theoretical electronic structure calculations of the bulk InVO$_4$ have also been calculated by first principle calculations within standard exchange-correlation functionals (reported a band gap of 3.1 eV).\cite{Oshikiri,Oshikiri11,PCCP} Apart from these, there were no theoretical studies exploring the high pressure behavior of InVO$_4$ to compliment the experiments. To the best of our knowledge, no theoretical or experimental studies on electronic and optical properties of the high pressure phase were reported. This kind of high pressure study of any material is very important in understanding their optoelectronic properties as the knowledge about optical properties of materials at high pressure provides insight about their performances in practical applications. With this motivation, in this study, we present theoretical calculations for structural properties in comparison with the experiment as well as the electronic and optical properties for both the phases of InVO$_4$.\\
The remaining paper is organized as follows. In section II, we briefly describe the crystal structure and computational details. Results and discussion concerning the structural, electronic, and optical properties of InVO$_4$ and the structural transformation are presented in section III, and finally, section IV summarizes the conclusions.

\section{Crystal Structure and Calculational details}
\begin{figure}
      \centering
       \includegraphics[width=5.5in]{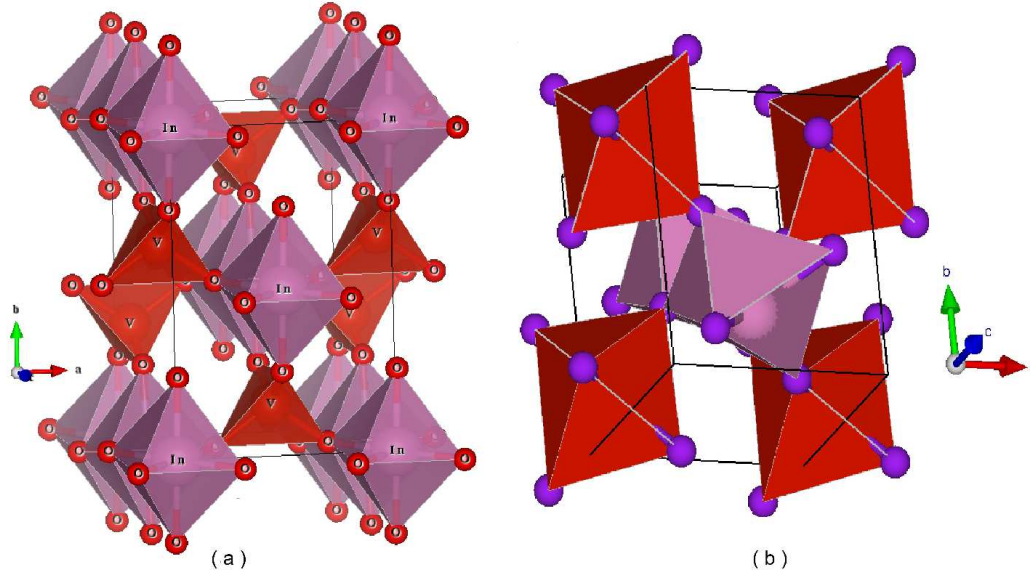}\\
       \caption{(color online) Crystal structure of (a) Orthorhombic and (b) Monoclinic InVO$_4$}
        \label{phase}
     \end{figure}

From previous reports, it was found that InVO$_4$ is a polymorphic material, which can be found in five different types of structures\cite{McCrone,Touboul1,Daniel} in nature. Among them, phase-I is monoclinic and is a temperature dependent phase. Phase-III is orthorhombic and found at ambient pressure and phase-V is monoclinic, which is the high pressure phase. From recent experiments,\cite{Daniel} a phase transition from phase-III (orthorhombic) to phase-V (monoclinic) at 7 GPa is found to occur. On the other hand,  phase-II and phase-IV are intermediate phases, and there structures are still unknown. In this work, we have only discussed about the crystal structure of phase-III, which is orthorhombic, and phase-V, which is monoclinic. The crystal structures of orthorhombic phase is  made up of $InO_6$ octahedra and $VO_4$ tetrahedra, and the $InO_6$ octahedra are connected by sharing the edges forming a linear chain of octahedra units along the \textit{c}-axis, while the chains are linked by $VO_4$ sharing only the corners. In the case of monoclinic structure, these blocks are $InO_6$ and $VO_6$ octahedra.\cite{Daniel} The crystal structure of orthorhombic and monoclinic phases of InVO$_4$ are shown in Fig. \ref{phase}.
The projector augmented wave (PAW) approach has been used to perform the \emph{ab-initio} calculations, which is implemented through vienna ab-initio simulation package\cite{Kresse} based on density functional theory (DFT).\cite{Hohenberg,Kohn} The generalized gradient approximation (GGA)\cite{Burke} was used to treat electron-electron interactions in InVO$_4$. The Broyden-Fletcher-Goldgarb-Shanno (BFGS) minimization scheme\cite{Almlof} has been used for structural relaxation. The convergence criteria for structural optimization of both the InVO$_4$ structures were performed with a kinetic energy cutoff of 900 eV and we have used Monkhorst Packing\cite{Monkhorst} as (9 9 8) for gamma centered grid. The valence electronic configurations of the atoms considered here are as follows: In: [5s$^2$5p$^1$], V: [3p$^6$3d$^4$4s$^1$], and O: [2s$^2$2p$^4$]. The self-consistent energy convergence was set to $1\times10^{-8}$ eV/atom. The convergence criterion for the maximal force between atoms was 0.01 eV/$\textup{\AA}$. For calculating the electronic structure, full potential linear augmented plane wave (FP-LAPW) method implemented using WIEN2k package has been used with the Perdew-Burke-Ernzerhof (PBE) and Tran-Blaha Modified Becke-Johnson (TB-mBJ) functional. The performance of TB-mBJ functional for calculating the band structure and the band gap of semiconductors and the insulators is more accurate. After getting convergence, the K-point is fixed to 5000 in order to obtain self consistency in irreducible Brillouin zone. In our calculation, we set the energy separation between core and valance electrons as -6 Ry. Radius of muffin tin for all the atoms are chosen as default value.

\section{Results and discussion}

\subsection{Structural properties of Polymorphic InVO$_4$}
    \begin{figure}
        \centering
       \includegraphics[width=4in]{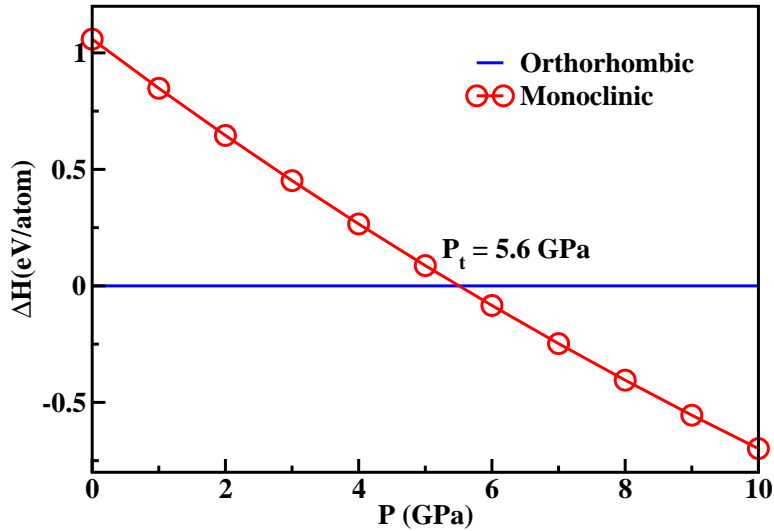}\\
       \caption{(color online) Phase transition from orthorhombic to monoclinic InVO$_4$ phase. Change in enthalpy per unit cell with respect to pressure.}
       \label{enthalpy}
    \end{figure}

As discussed in Section II, the polymorphic InVO$_4$ crystallizes in orthorhombic structure (with space group \textit{Cmcm}) at ambient conditions and transforms to monoclinic structure (with space group \textit{P2/c)} under compression. As a first step, we have performed geometry optimization with full structural relaxations including lattice parameters and atomic positions to compare the experimentally reported transition. Here, the theoretical transition pressure in polymorphic InVO$_4$ can be clearly depicted from the hydrostatic pressure dependance of enthalpy from 0 to 14 GPa with a step size of 1 GPa. For instance, we included the differences in enthalpies for orthorhombic and monoclinic structures and are shown in Fig.\ref{enthalpy}. From this, it can be clearly seen that the orthorhombic phase is stable at ambient conditions (since enthalpy less than other structure) and monoclinic structure is energetically favourable from above 6 GPa. Note that our calculated transition pressure (5.6 GPa) agrees well with the experimental value of 7 GPa. Further, we calculated ground state and high pressure structural parameters (lattice parameters, atomic positions, and bond lengths) of both the structures and are compared with those of the experimental data, as shown in Tables \ref{I}-\ref{III}. From table \ref{I}, the calculated volumes are  slightly overestimated with the experimental data, which is due to the choice of functional used in the current study. The other parameters such as atomic positions and bond lengths calculated agree quite well with the experiments.

\begin{table}
\caption{Calculated structural properties of orthorhombic and monoclinic InVO$_4$ at 0 GPa and 6 GPa respectively using PAW method with the experimental data (see in Ref. \onlinecite{Daniel}). a, b and c are the lattice parameters (in \text{\AA}), V is the volume (in \text{\AA}$^3$) of the unit cell, $B_0$ and $B'_0$ are bulk modulus (in GPa) and its pressure derivatives.}
\begin{ruledtabular}
\begin{tabular}{cccccccc}
 & \multicolumn{2}{c}{Ortho} & & \multicolumn{2}{c}{Mono}  \\
 \hline
Parameter &  Theo.  & Exp. & & Theo. & Exp. \\ \hline
a & 5.816 & 5.738 & & 4.776 &4.714  \\
b & 8.739 & 8.492 & &5.588  & 5.459 \\
c &6.775  & 6.582 & & 4.958 & 4.903 \\
$\beta$& 90 & 90  & &92.89  &93.8  \\
V &344.37  & 320.72 & &132.17 & 125.89  \\
B$_0$  &76.47  & 69 & &183  &168  \\
$B_0^\prime$ & 3 &4  & & 6 &4  \\
\end{tabular}
\end{ruledtabular}
\label{I}
\end{table}

\begin{table}
\caption{Atomic positions of orthorhombic and monoclinic InVO$_4$ at 0 GPa and 6 GPa along with the experiments data in Ref. \onlinecite{Daniel}.}
\begin{ruledtabular}
\begin{tabular}{cccccccc}
 & \multicolumn{2}{c}{Ortho} & & \multicolumn{2}{c}{Mono}  \\ \hline
Parameter &  Theo.  & Exp. & & Theo. & Exp. \\ \hline
In(4a)&(0, 0, 0)&(0, 0, 0)&In(2f)&(0.5, 0.704, 0.25)&(0.5, 0.711, 0.25)\\
V(4c)&(0, 0.358, 0.25)&(0, 0.3617, 0.25)&V(2e)&(0, 0.181, 0.25)&(0, 0.159, 0.25)\\
$O_1$(8g)&(0.256,0.473,0.25)&(0.2568,0.483,0.25)&$O_1$(4g)&(0.21,0.91,0.46)&(0.214,0.861,0.492)\\
$O_2$(8f)&(0,0.753,0.953)&(0,0.7492,0.9573)&$O_2$(4g)&(0.248,0.382,0.386)&(0.242,0.407,0.399)\\
\end{tabular}
\end{ruledtabular}
\label{II}
\end{table}

\begin{table}
\caption{Calculated bond lengths for the orthorhombic and monoclinic InVO$_4$ at 0 GPa and 6 GPa along with the experiments (see in Ref. \onlinecite{Daniel}). In-$O_2$($\times2$) indicates two equal bonds of same length between In and O$_2$ atoms ( in \text{\AA}) and In-$O_1$($\times4$) indicates four equal bonds of same length between In and O$_1$ atoms.}
\begin{ruledtabular}
\begin{tabular}{cccccccc}
 & \multicolumn{2}{c}{Ortho} & & \multicolumn{2}{c}{Mono}  \\ \hline
Parameter &  Theo.  & Exp.
 & & Theo. & Exp. \\ \hline
In-$O_2$($\times2$) &2.22  & 2.1483 & In-$O_1$($\times2$) & 2.11 &2.0268 \\
In-$O_1$($\times4$) &2.18  & 2.1623 & In-$O_2$($\times2$) &2.18  &2.1397 \\
 & & & In-$O_2$($\times2$)& 2.29 &  2.2101\\
V-$O_2$($\times2$) &1.68  & 1.6579 & V-$O_1$($\times2$) & 1.75 & 1.673\\
V-$O_1$($\times2$) &1.79  & 1.7983 & V-$O_1$($\times2$) & 2.08 & 2.2166\\
  & & & V-$O_2$($\times2$) &1.87 &1.8861 \\
\end{tabular}
\end{ruledtabular}
\label{III}
\end{table}

\begin{figure}
         \centering
        \includegraphics[width=6in]{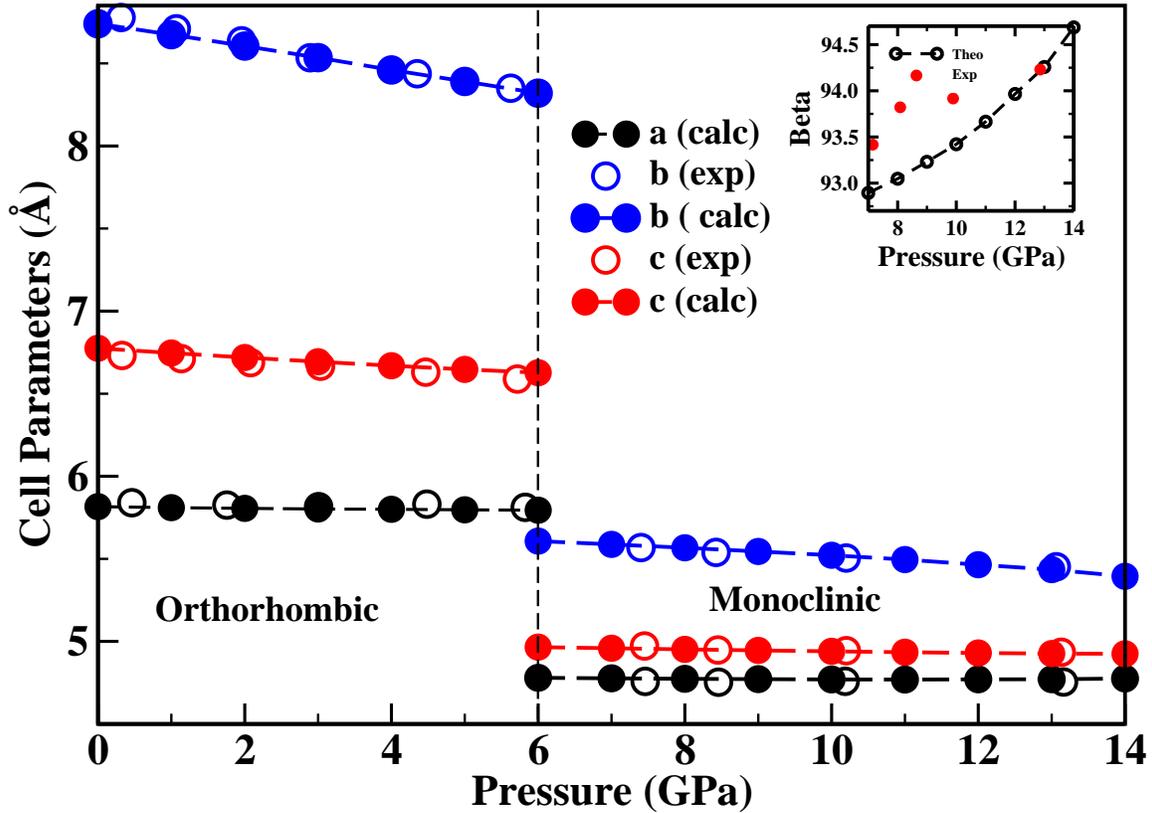}\\
        \caption{(color online) Pressure dependence of lattice parameters $a$ (black), $b$ (blue), and $c$ (red). The closed symbols with dashed line and the open symbols respectively denote the theoretical result and the experimental result\cite{Daniel}.}
        \label{3}
     \end{figure}

Now, we move to analyze the effect of hydrostatic pressure on lattice parameters and volumes of both structures. The variation in the lattice parameters (\textit{a}, \textit{b}, and \textit{c}) with respect to pressure along with experiment is shown in Fig. \ref {3}. From this plot, it is clear that the lattice parameters are decreasing with pressure, and it can be seen that there is a sudden change in the obtained parameters, which is expected due to structural transition. It is also clear from the Fig. \ref {3} that the calculated values are in good agreement throughout the pressure range with the experimental values. The compressibility of each lattice parameter according to pressure variation can be well described with the first order pressure coefficients. The calculated first order pressure coefficients for lattice parameters \textit{a}, \textit{b}, and \textit{c} of orthorhombic phase are 0.35$\times 10^{-2}$ $\AA$/GPa, 7.04$\times 10^{-2}$ $\AA$/GPa, and 2.47$\times 10^{-2}$ $\AA$/GPa, respectively. For monoclinic phase, these values are 0.03$\times 10^{-2}$, 2.72$\times 10^{-2}$, and 0.49$\times 10^{-2}$ $\AA$/GPa, respectively. From these values, it can be said that the first order pressure coefficient of lattice parameter b $\textgreater$ c $\textgreater$ a for both the phases. This also shows that the lattice parameter \textit{b} is the most compressible. This can be expected as from the arrangement of the InO$_6$ and VO$_4$ polyhedra in the unit cell. The anisotropic compression can be related with the fact that there are no VO$_4$ tetrahedra in between the InO$_6$ octahedra along the \textit{b}-axis direction.

\begin{figure}
  \centering
  \includegraphics[width=6in]{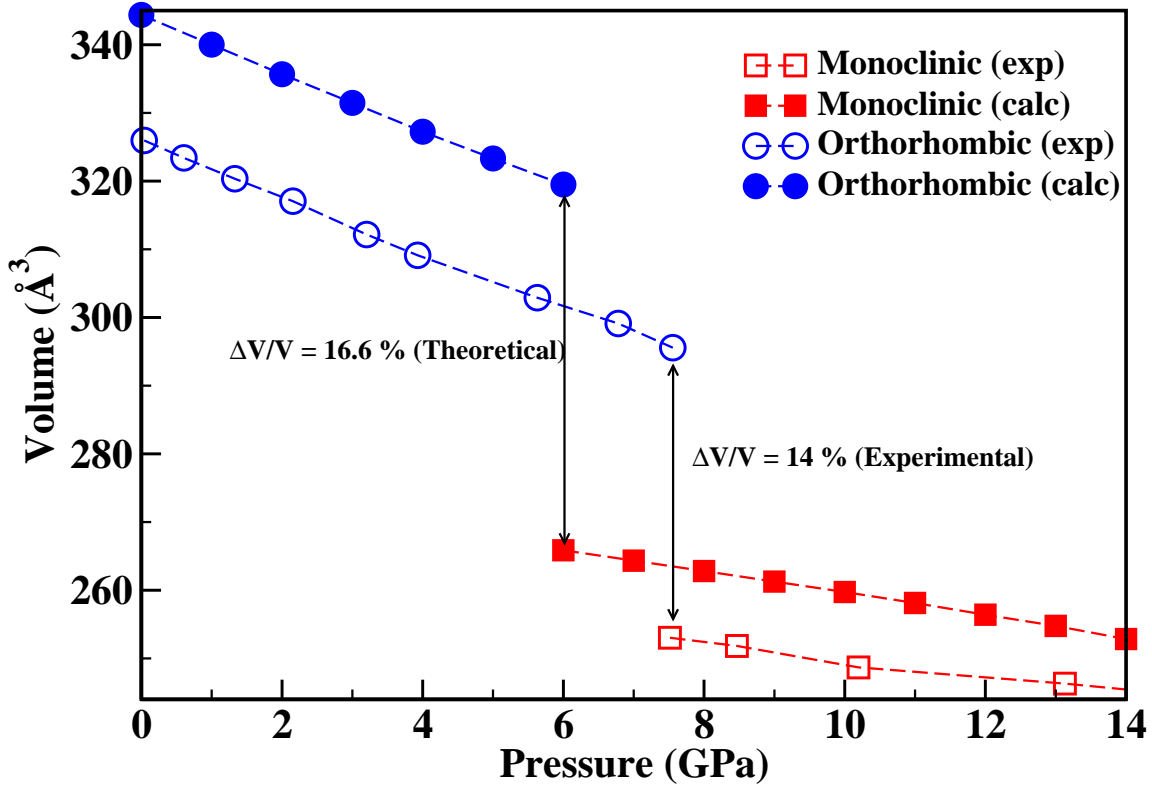} \\
  \caption{(color online) Calculated pressure dependence of the crystal volume for InVO$_4$ polymorphic phases along with experiments\cite{Daniel}.}
  \label{4}
 \end{figure}
\begin{figure}
  \centering
  \includegraphics[width=6in]{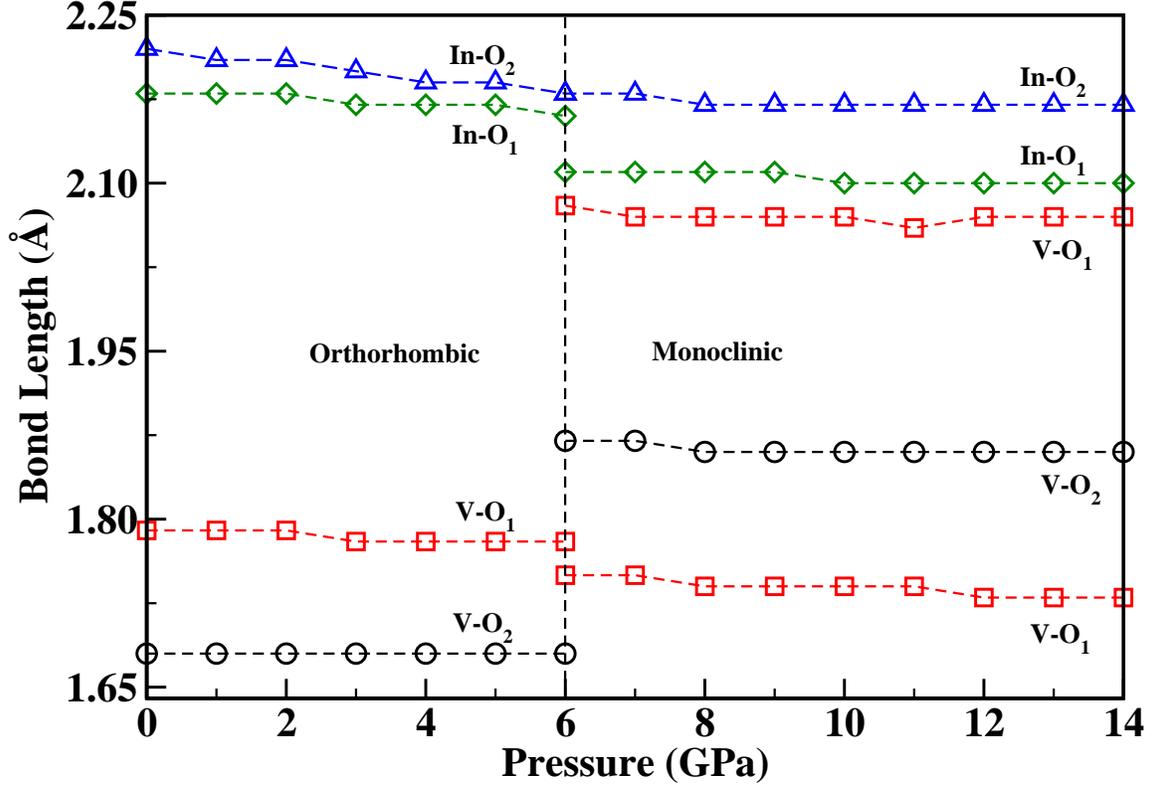}
  \caption{(color online) Calculated pressure dependence of bond lengths for InVO$_4$ polymorphic phases.}
  \label{5}
  \end{figure}

 In addition, we have also plotted volume as a function of pressure, which is shown in the Fig. \ref {4}. From this plot, it can be seen that there is a large volume collapse of nearly 16.6$\%$, which is close to the experimental value of 14 $\%$. This confirms the first-order character of the phase transition, which is taking place under the application of pressure from CrVO$_4$ type orthorhombic InVO$_4$ to wolframite type monoclinic InVO$_4$ structure. As depicted in Fig. \ref{4}, ambient phase orthorhombic structure is more compressible when compared to high pressure phase of monoclinic structure. On the other hand, from the obtained P-V data, we can get the equation of state (EOS) parameters using Murnaghan equation,\cite{murnaghan} which are given in the Table \ref{I} along with the experimental values. From the table \ref{I}, we can see that the calculated values are in good agreement with experimental values, where B$_0$ is the bulk modulus and $B_0^\prime$ is its pressure derivative. From this analysis, it can be said that the bulk modulus for the monoclinic structure is nearly three times larger than that of the low pressure orthorhombic structure. This is consistent with large volume collapse associated with the phase transition and the fact that the monoclinic phase is much denser than that of the orthorhombic phase. On the other hand, this indicates the increase in coordination number around vanadium atom. Note that the computed bulk modulus for CrVO$_4$ type InVO$_4$ is also comparable with previously reported results of InPO$_4$.\cite{Moreno} The reported bulk modulus (B$_0$) and its pressure derivative ($B_0^\prime$) for InPO$_4$ is 78.65 GPa and 4.05, which are of the same order for InVO$_4$ given in Table \ref{I}. In addition, we also computed the bond lengths as a function of pressure for both structures and are shown in Fig. \ref {5}. From this, we found that all the bond lengths are decreasing with pressure and also the In-O$_2$ bond length in orthorhombic structure is more compressible than the other.

\subsection{Electronic Structure}
\begin{figure}
\subfigure[]{\includegraphics[width=3in,clip]{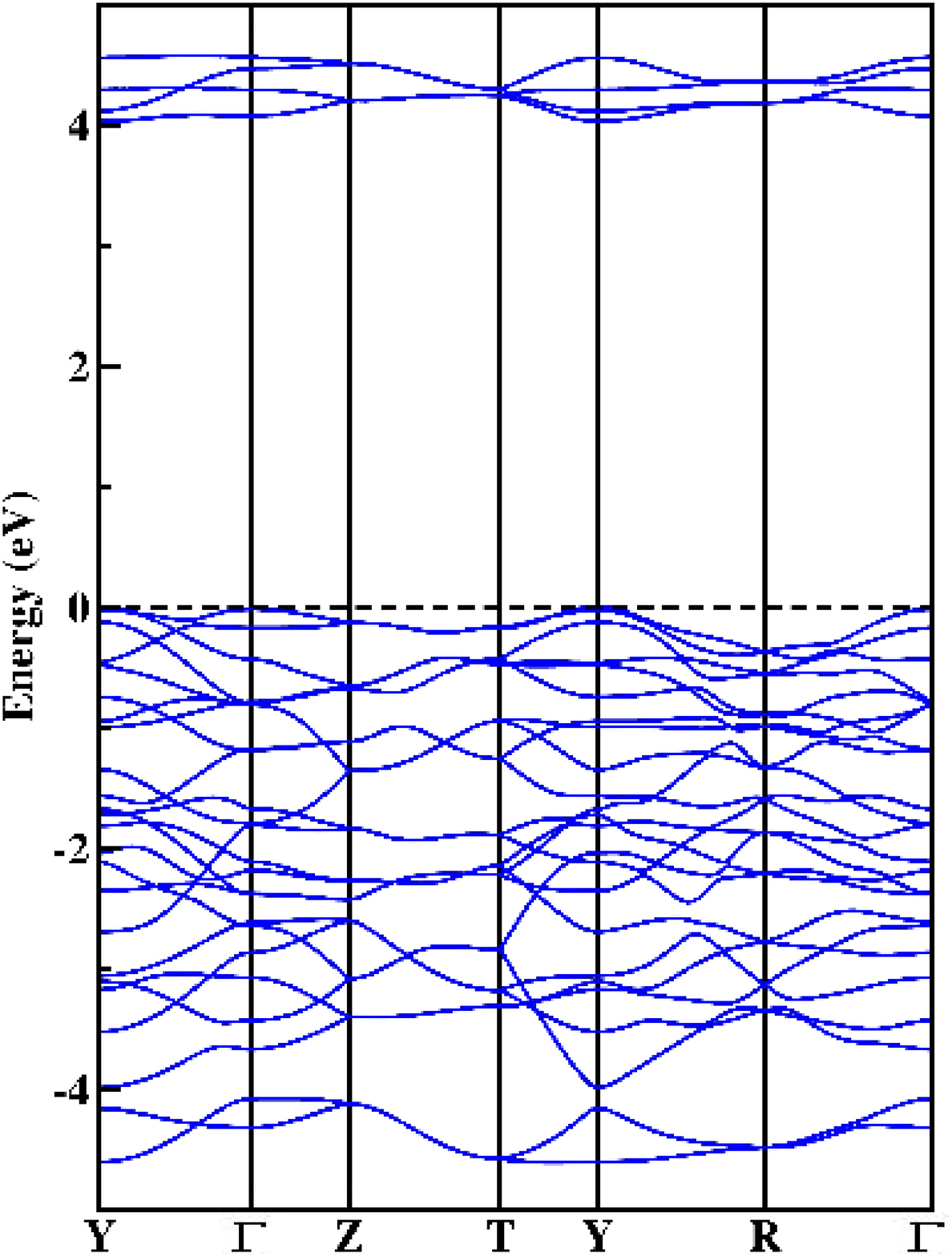}}
\subfigure[]{\includegraphics[width=3in,clip]{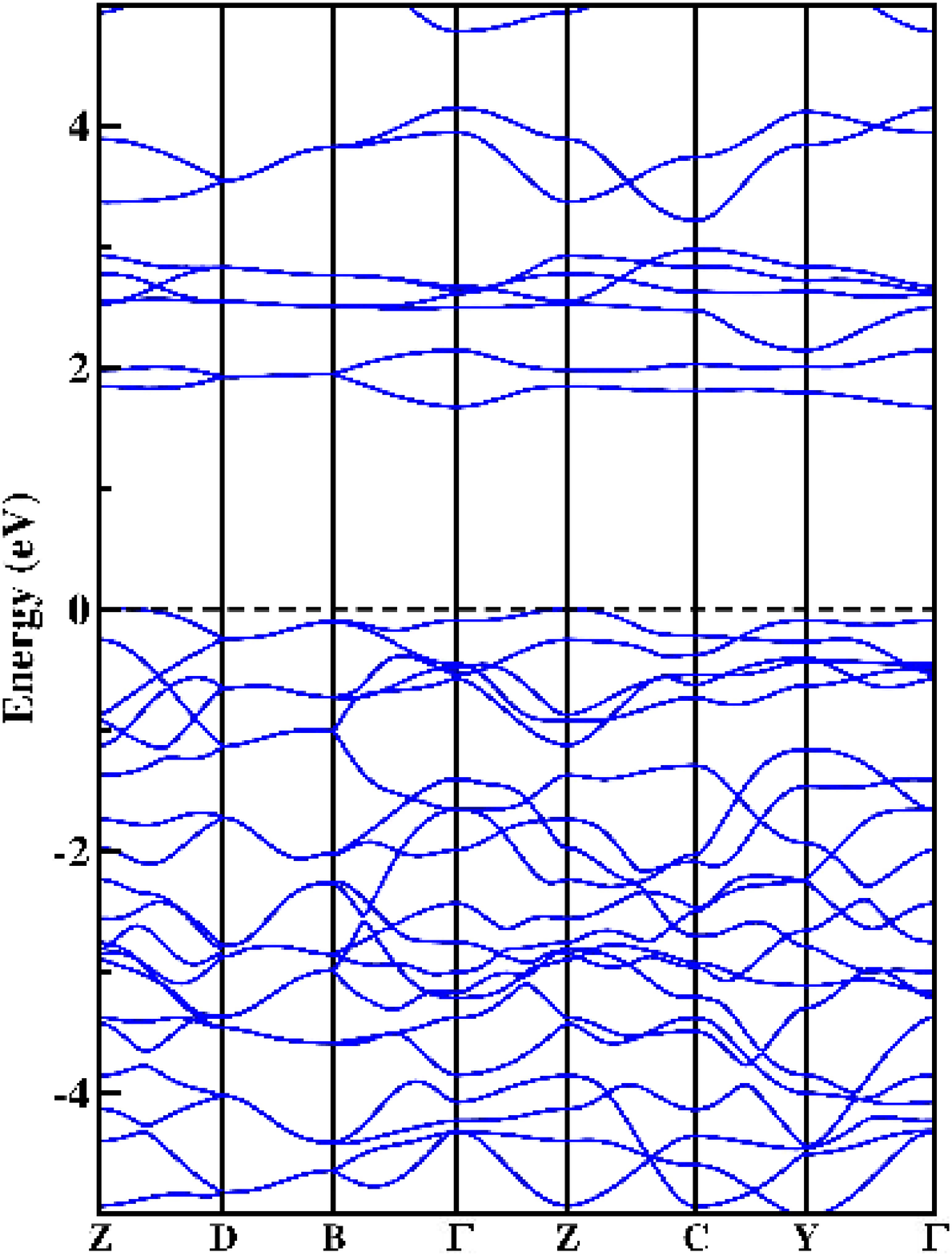}}
\caption{(color online) Calculated band structure of (a)orthorhombic and (b)monoclinic phases of InVO$_4$ at the optimized lattice parameter. The zero energy is taken at E$_f$=0.}
\label{6}
\end{figure}

\begin{figure}
\begin{tabular}{c}
\subfigure[]{ \includegraphics[width=4.5in]{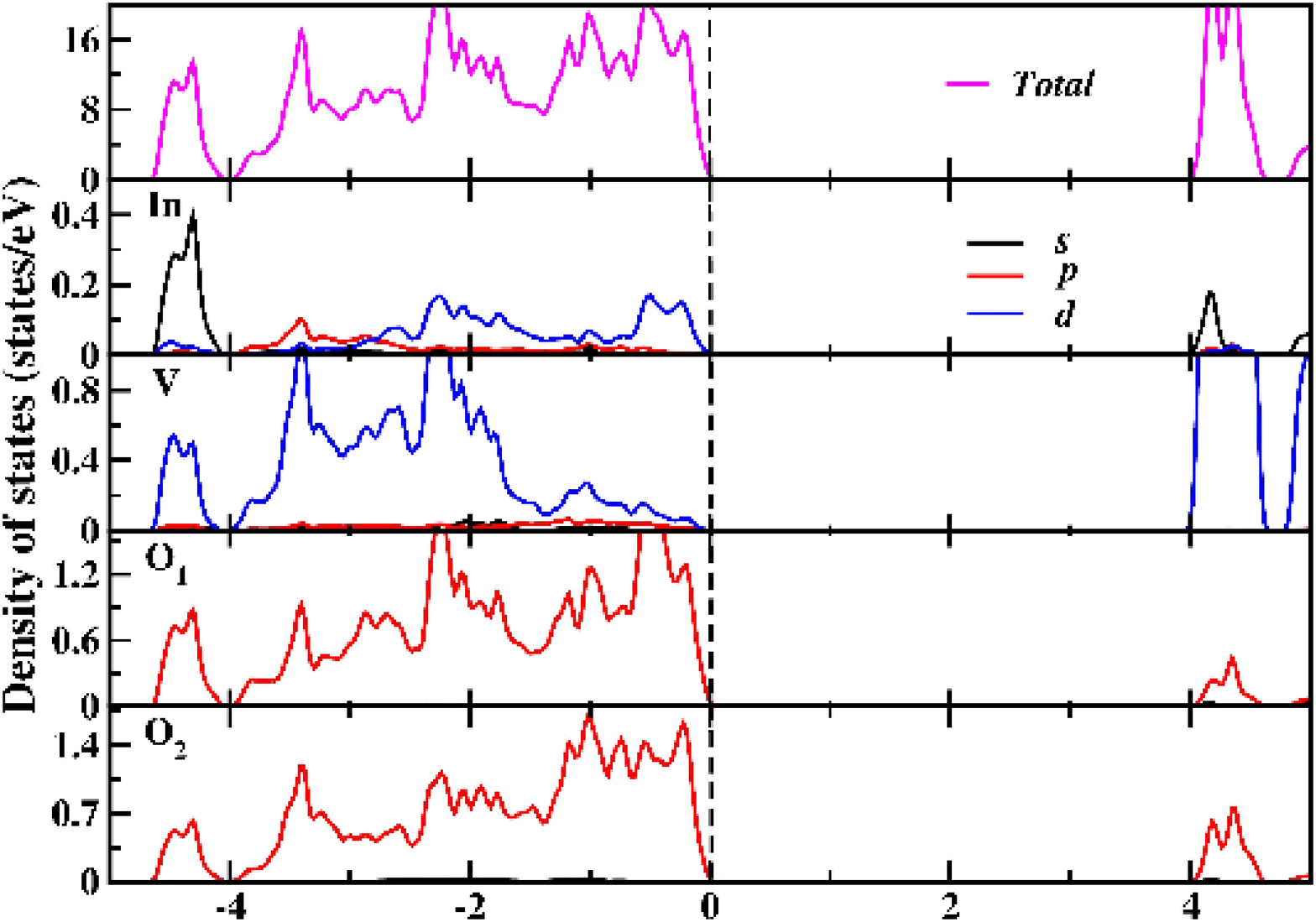}}
\\
\subfigure[]{ \includegraphics[width=4.5in]{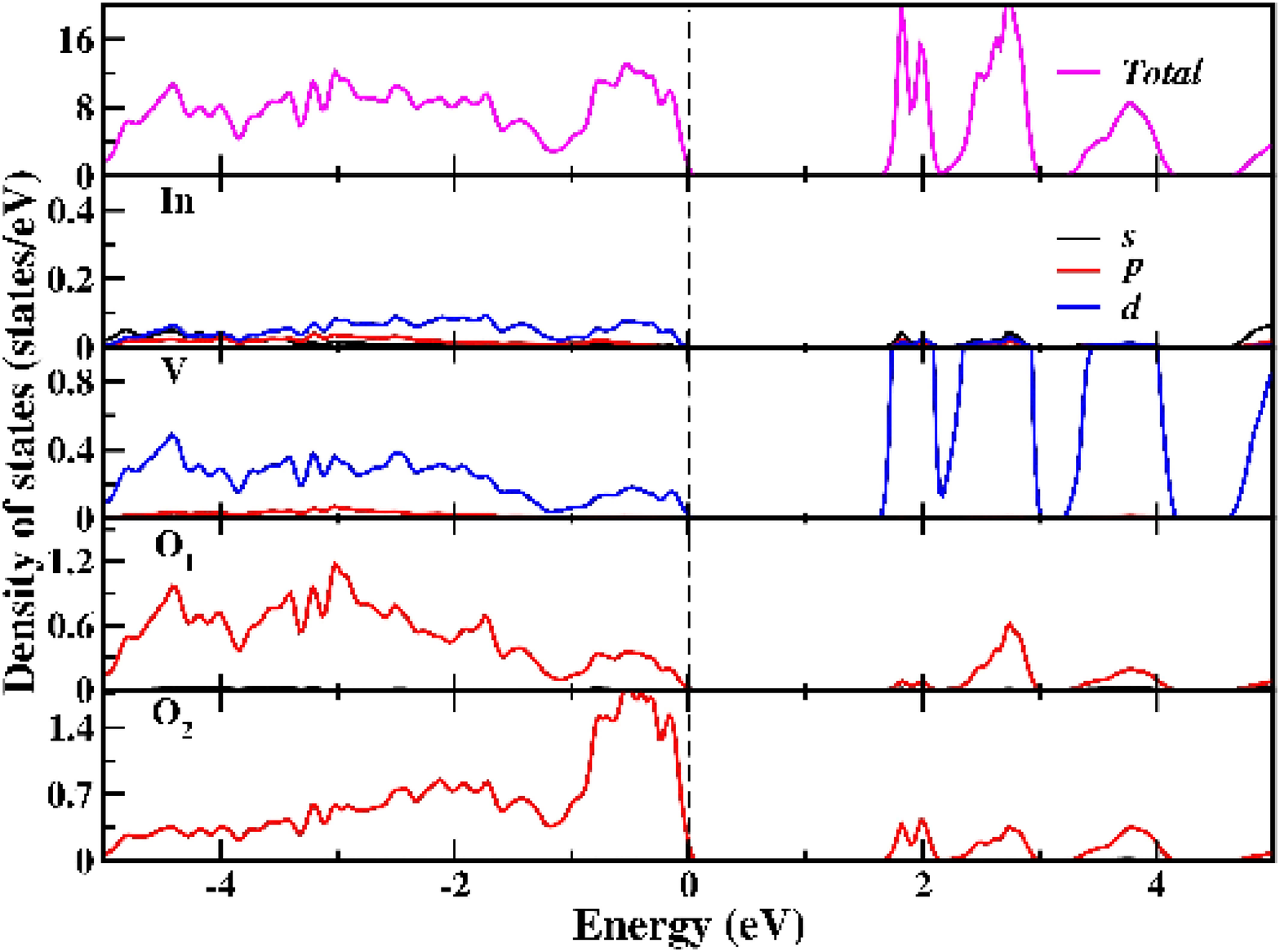}}
\end{tabular}
\caption{(color online) Density of states of (a)orthorhombic and (b)monoclinic phases of InVO$_4$. The zero energy is taken at E$_f$.}
\label{7}
\end{figure}
The calculated band structures of the orthorhombic and monoclinic phases of InVO$_4$, using the FP-LAPW approach within TB-mBJ functional along the various symmetry lines, are shown in Fig. \ref {6}. These calculations were performed at optimized structures as discussed in Section III A. It is a known fact that the standard DFT functional local density approximation (LDA)/GGA always underestimate band gap.\cite{Singh,Tran,Koller,Dixit,Camargo} In order to compare, we have used GGA functional along with the TB-mBJ functional for calculating band gaps. The obtained band gap for the orthorhombic phase is 3.13 eV using PBE method and 4.02 eV using TB-mBJ functional. For the monoclinic structure, the calculated band gaps are 1.67 and 1.04 eV using TB-mBJ and PBE functional, respectively. The calculated band gap from GGA functional for orthorhombic structure is in good agreement with the previously reported theoretical band gap of nearly 3 eV\cite{Oshikiri,Oshikiri11,PCCP} using LDA/GGA functional. It is to be noted that earlier experiments based on optical absorption measurements indicate 2 eV band gap for orthorhombic InVO$_4$.\cite{V2, Yea} The difference between the presently computed band gap and experimental band gap for orthorhombic phase might be due to the electron excitonic effect, which is beyond the scope of TB-mBJ functional calculation.

\begin{figure}
      \centering
       \includegraphics[width=6in]{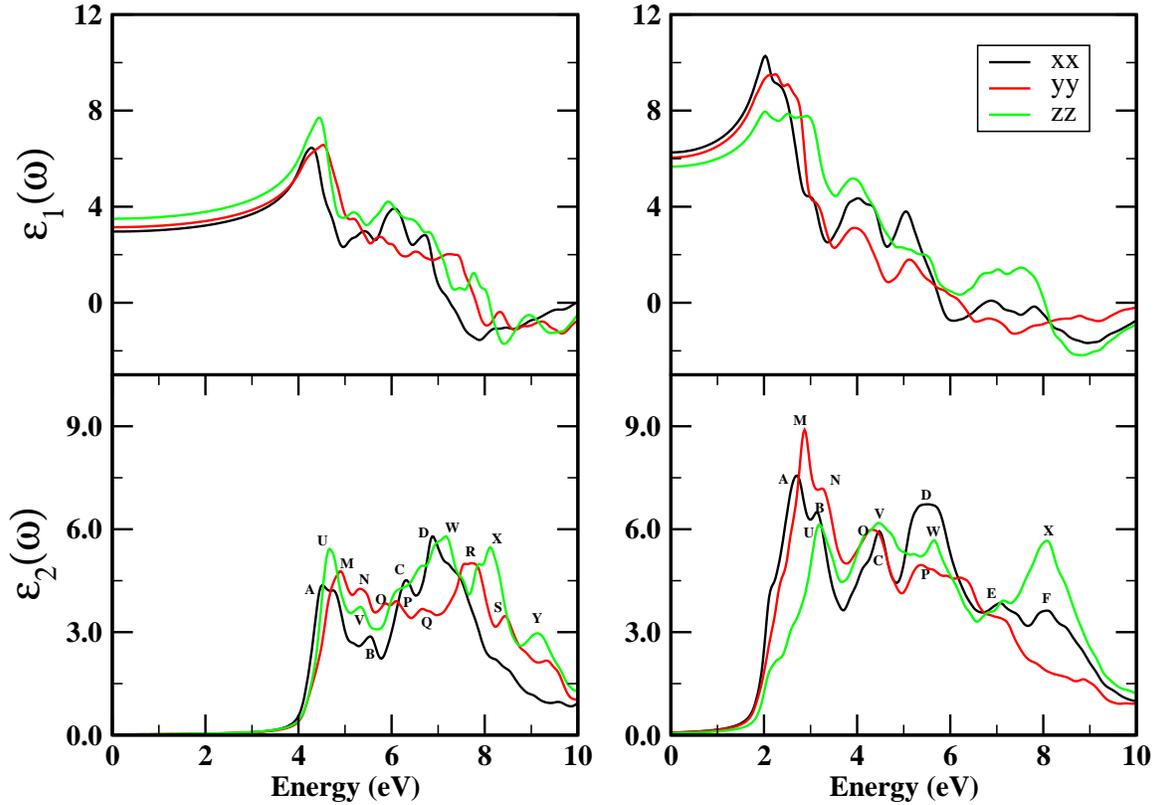}\\
       \caption{(color online) Calculated complex dielectric function of orthorhombic and monoclinic InVO$_4$ using TB-mBJ functional. The real part of this function is shown in the above and the imaginary part is shown in the below. Left side of this figure indicates the orthorhombic and the right side indicates the monoclinic phase of InVO$_4$.}
        \label{8}
     \end{figure}
From the electronic band structure shown in the Fig. \ref {6}, it is found that the orthorhombic InVO$_4$ is a direct band gap material as the band gap occurs along the Y direction between the valence band maximum (VBM) and conduction band minimum(CBM). The monoclinic InVO$_4$ is an indirect band gap material as the band gap occurs along the $\Gamma$-Z direction between the VBM and CBM. Except this indirect band gap, there are two more minimum direct band gaps present in Z and $\Gamma$ directions, which are 1.78 and 1.84 eV, respectively. As these direct band gaps are slightly larger in magnitude than the indirect band gap, they are preferred for optical transitions and can contribute in the visible light absorption.
Fig. \ref {7} shows the density of states(DOS) and partial density of states (PDOS) of both the structures, which have been calculated using TB-mBJ functional. It can be clearly seen from the partial density of states(PDOS) of the orthorhombic InVO$_4$ that between -5 eV to -3.5 eV, there are mainly two peaks available because of the \textit{s}-states of the indium, \textit{d}-states of the vanadium, and the \textit{p}-states of the oxygen. The states, predominant between -3.5 and -2 eV, are mainly because of vanadium atom, and this contribution is mainly due to \textit{d}-states. Near the fermi level between -2 to 0 eV, the main contribution is from the oxygen \textit{p}-states. So, it is clear that the top of the valence band is composed mainly of \textit{p}-states of oxygen. It is also clear from the Fig. \ref {7} that the bottom of the conduction band is mainly composed of \textit{d}-states of vanadium atom including little contribution due to \textit{s}-state of indium and \textit{p}-state of oxygen at about 4 eV. On the other hand, for monoclinic InVO$_4$, there are very less contributions due to indium atom. Vanadium atom dominates through a wide range (-4.5eV to -2.5eV) due to \textit{d}-states. From -4.5 to -3 eV, the main contribution is from the \textit{p}-states of $O_1$ atom. At nearly -3.5 eV, there is an overlap between the vanadium atom and the $O_1$ atom, which suggests a strong covalent behavior between vanadium and oxygen atoms within the ($[VO4]^{3-}$). The \textit{2p}-states of $O_2$ are dominated in between -1 to 0 eV, which confirms that the top of the valance band is mainly due to \textit{p}-states of the oxygen atom. The bottom of the conduction band is mainly composed of \textit{d}-states of vanadium atom with a little contribution of \textit{p}-states of $O_2$ atom at about 1.7eV.
\begin{figure}
      \centering
       \includegraphics[width=6in]{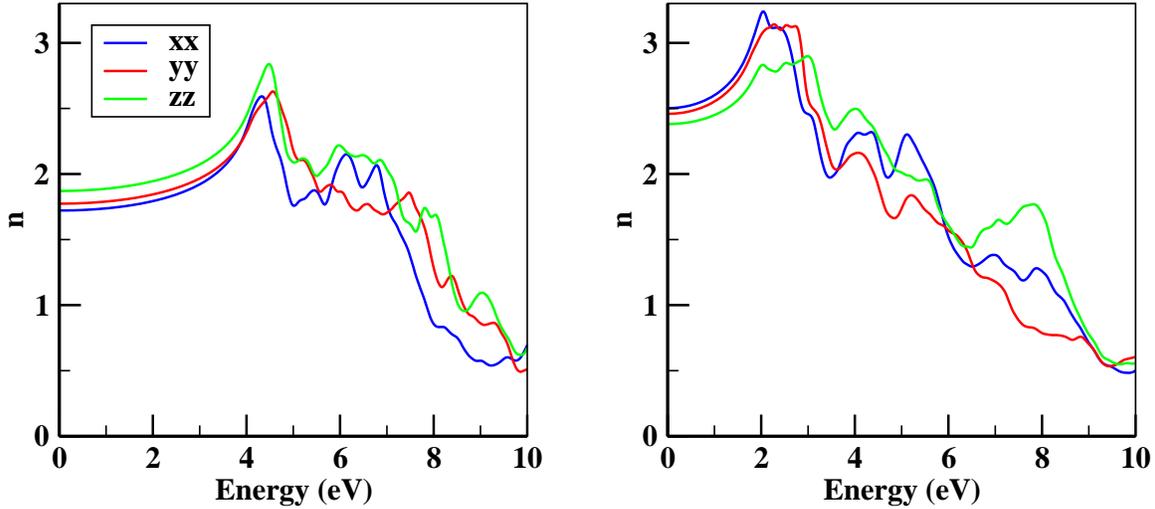}\\
       \caption{(color online) Index of refraction of orthorhombic (left) and monoclinic (right) InVO$_4$}
        \label{9}
     \end{figure}
It can be seen from this discussion that for both the structures, the conduction band minima mainly consists of \textit{d}-states of vanadium atom and the valence band maximum mainly consists of \textit{p}-states of oxygen atom, favoring both the structures as photo-catalysts.\cite{Oshikiri,Oshikiri11} In comparison to other photocatalytic materials, such as \textit{4d} and \textit{5d} metal compounds InNbO$_4$ and InTaO$_4$ and the well known photocatalyst TiO$_2$, orthorhombic InVO$_4$ (\textit{3d}) has already proven to be a good photocatalyst for wider wave length region. The band gaps are 3.7 eV (InTaO$_4$), 3.4 eV (InNbO$_4$), 3.1 eV (InVO$_4$), and 2.5 eV (TiO$_2$), respectively.\cite{Oshikiri,Oshikiri11} It has been shown that in order to active a photocatalyst in the longer wavelength region, one needs to lower the bottom of the \textit{3d} conduction band. This help us to conclude that the monoclinic phase could be a promising photo-catalyst than the orthorhombic phase, as the bottom of the conduction band is lower than the orthorhombic phase.
\subsection{Optical Properties}
To calculate the optical properties for the orthorhombic and monoclinic phase of InVO$_4$, we have used the TB-mBJ functional. Optical properties of any material can be well described by its dielectric function, which is given by $\epsilon(q, \omega)=\epsilon_1(\omega)+\imath\epsilon_2(\omega)$, where q is the momentum transfer in the photon-electron interaction and $\omega$ is the energy transfer. The imaginary part $\epsilon_2(\omega)$ of the dielectric function $\epsilon(q, \omega)$ represents the direct inter band transitions between valence band and conduction band states. Here, in this calculation, we have neglected the intraband transitions as contribution from this type of transition is very less. From the selection rule, it can be said that the inter band transitions occur only in momentum quantum number states given by $\Delta$L=$\pm$1. The real part of the dielectric tensor $\epsilon_1$ can be calculated using the Kramer's-Kronig transformation. Here, in this study, $\epsilon_2$ has been calculated up to 10 eV.
\begin{table}[ht]\centering
 \caption{The observed optical transitions of orthorhombic and monoclinic InVO$_4$ from imaginary
part of dielectric function using TB-mBJ electronic structure.}
 \begin{center}
    \begin{tabular}{ccccccc}
      \hline
       & \multicolumn{2}{c}{Along xx} & \multicolumn{2}{c}{Along yy} & \multicolumn{2}{c}{Along zz} \\
      Orthorhombic & A & 4.54  & M & 4.95 & U & 4.7 \\
      InVO$_4$  & B & 5.6 & N & 5.35 & V & 5.35 \\
        & C & 6.3 & O & 5.9 & W & 7.2 \\
        & D & 6.9 & P & 6.14 & X & 8.15 \\
        & - & - & Q & 6.7 & Y & 9.12 \\
        & - & - & R & 7.7 & - & -  \\
        & - & - & S & 8.4 & - & - \\
      Monoclinic & A & 2.74 & M & 2.9 & U & 3.2 \\
      InVO$_4$ & B & 3.15 & N & 3.3 & V & 4.4 \\
       & C & 4.5 & O & 4.3 & W & 5.6 \\
       & D & 5.5 & P & 6.4 & X & 8 \\
       & E & 7 & - & - & - & - \\
       & F & 8 & - & - & - & - \\
      \hline
    \end{tabular}
   \end{center}
   \label{IV}
\end{table}
\begin{figure}
      \centering
       \includegraphics[width=6in]{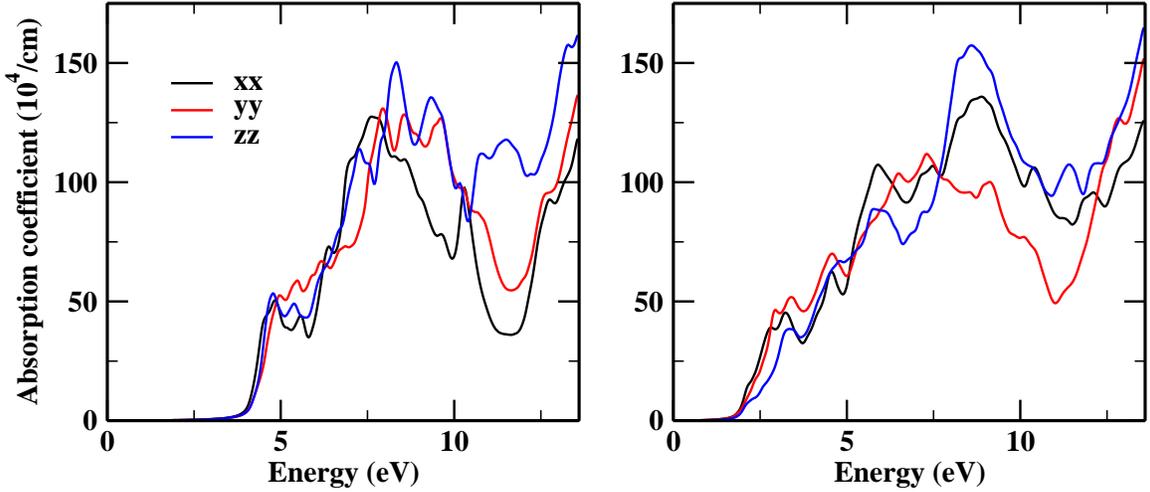}\\
       \caption{(color online) Absorption spectra of both the orthorhombic (left) and monoclinic (right) InVO$_4$ using TB-mBJ functional.}
        \label{10}
     \end{figure}
The important optical properties of any material such as absorption and refractive index can be calculated from the proper knowledge of the real and imaginary parts of the dielectric function. Fig. \ref {8} shows the real and imaginary part of the dielectric function of orthorhombic and monoclinic InVO$_4$. In this figure, A, B, C, D, E, and F along the \textit{XX} direction, M, N, O, P, Q, R, and S along \textit{YY}, and U, V, W, X, and Y along the \textit{ZZ} direction are the prominent peaks. As the peaks in the imaginary part of the dielectric function mainly arise due to the inter-band transitions between VBM and CBM in the different energy levels, using these peaks of imaginary part of the dielectric function and the calculated TB-mBJ electronic band structure shown in Fig. \ref {6}, we have calculated different inter-band transitions. In case of Orthorhombic InVO$_4$, the allowed transitions are as follows: (a) O(p)$\rightarrow$V(d) in the 4 to 6 eV energy range, (b) V(d)$\rightarrow$O(p) in the 6-8 eV range, (c) In(s)$\rightarrow$O(p) in the 8-9 eV range. Among these transitions, the dominant part is the O(p)$\rightarrow$V(d). In the case of monoclinic InVO$_4$, the allowed transitions are (a) O(p)$\rightarrow$V(d) in the 1.67-3.5 eV range, (b) O(p)$\rightarrow$V(d) in the 3.5-5.5 eV range, and (c)V(d)$\rightarrow$O(p) in the 2-4 eV range. In both cases, the dominant transitions are taking place in the same manner except the decrease in the band gap values for monoclinic InVO$_4$. These allowed transitions are given in the Table \ref{IV}. The calculated values of real part of the dielectric constants near to zero photon energy are $\epsilon_{1}^{XX}(0)=2.96$, $\epsilon_{1}^{YY}(0)=3.14$, and $\epsilon_{1}^{ZZ}(0)=3.49$ for orthorhombic phase and $\epsilon_{1}^{XX}(0)=6.26$, $\epsilon_{1}^{YY}(0)=6.05$, and $\epsilon_{1}^{ZZ}(0)=5.67$ for monoclinic phase. Refractive index corresponding to this photon energy is $n_{XX}(0)=1.72$, $n_{YY}(0)=1.77$, and $n_{ZZ}(0)=1.87$ for orthorhombic phase and $n_{XX}(0)=2.5$, $n_{YY}(0)=2.46$, and $n_{ZZ}(0)=2.38$ for monoclinic phase. It can be seen from the values given above that the index of refraction for orthorhombic phase is high in the ZZ direction and low in the XX direction at nearly 0 eV energy. In case of monoclinic phase, this value of refractive index is high in the XX direction and low in the ZZ direction. The calculated index of refraction in three different directions is shown in Fig. \ref {9}. The maximum indices of refraction shown in Fig. \ref {9} are $n_{XX}=2.57$ at 4.35 eV, $n_{YY}=2.61$ at 4.58 eV, and $n_{ZZ}=2.82$ at 4.5 eV for orthorhombic phase and $n_{XX}=3.23$ at 2.07 eV, $n_{YY}=3.13$ at 2.3 eV, and $n_{ZZ}=2.88$ at 3.03 eV for monoclinic phase. Fig. \ref {10} shows the absorption spectra of both the orthorhombic and monoclinic InVO$_4$ phases. It is clearly seen from the figure that both the structures show optical anisotropy in the three different directions and the optical properties are different for different directions. Also, the peaks for orthorhombic phase located approximately above 4 eV (230-250 nm) and at nearly 3 eV (380-450 nm) for monoclinic phase. The absorption peaks of monoclinic phase fall more under the visible light region than those of the orthorhombic phase. This indicates that the absorption in monoclinic InVO$_4$ is similar to the previously reported experimental values,\cite{V2, Yea} which helps up to conclude that the monoclinic phase has the ability to utilize the solar energy in wide range. This absorption of visible light by monoclinic phase of InVO$_4$ may be derived from the large and delocalized DOS around the CBM and small energy band gaps of the two direct optical transition points along Z and G as mentioned in the electronic structure discussion.

\section{Conclusions}
In summary, we have reproduced the structural phase transition of InVO$_4$ and found that the calculated transition pressure and structural parameters related to both structures are in good agrement with the previously reported  experimental data. The computed volume collapse of 16.6$\%$ across the transition pressure and a drastic increase in bulk modului are consistent with the experiments. The changes in the crystal symmetry from CrVO$_4$ type to wolframite type structure lead to change in the coordination number of vanadium atom from four to six. The hydrostatic pressure dependance structural properties show, both the phases are highly anisotropic, orthorhombic structure is more compressible than that of the monoclinic structure, and \emph{b}-axis is more compressible than other axes. Furthermore, the examined electronic band structure calculations indicate that the obtained band gap for monoclinic structure is found to be less than orthorhombic structure. Finally, we have calculated the optical properties for both structures and found considerable optical anisotropy. The obtained optical transitions for both structures are similar in nature except the lower value of the electronic band gap for monoclinic structure. As the orthorhombic InVO$_4$ is already a well established photo-catalyst, we can say, that monoclinic phase is more suitable for the photo-catalyst than the orthorhombic phase in the visible light range.
\section{Acknowledgments}
S. M. would like to thank school of physics, University of Hyderabad, for financial support during M.Tech. S. A  would also like to thank DRDO, India, through ACRHEM for financial support. Authors are also thankful to CMSD, University of Hyderabad, for providing computational facilities. \\

\end{document}